\documentclass[10pt, paper=a4, UKenglish]{article}
\usepackage{graphicx}
\usepackage[utf8]{inputenc}
\usepackage{etoolbox,xspace}

% ...
\inputencoding{latin1}
% ...
\inputencoding{utf8}
% ...
\def\Title#1{\begin{center} {\Large #1 } \end{center}}
\def\Author#1{\begin{center}{ \sc #1} \end{center}}
\def\Address#1{\begin{center}{ \it #1} \end{center}}

\newcommand\pubblock{\rightline{\begin{tabular}{l} Proceedings of CTD 2020\\ \pubnumber\\
         \pubdate  \end{tabular}}}

\newenvironment{Abstract}{\begin{quotation} \begin{center} 
             \large ABSTRACT \end{center}\bigskip 
   \begin{center}\begin{large}}{\end{large}\end{center}\end{quotation}}

\newenvironment{Presented}{\begin{quotation} \begin{center} 
             PRESENTED AT\end{center}\bigskip 
      \begin{center}\begin{large}}{\end{large}\end{center} \end{quotation}}

\def\Acknowledgements{\bigskip  \bigskip \begin{center} \begin{large}
      \bf ACKNOWLEDGEMENTS \end{large}\end{center}}

\def\beq{\begin{equation}}
\def\eeq#1{\label{#1}\end{equation}}
\def\eeqn{\end{equation}}

\def\beqa{\begin{eqnarray}}
\def\eeqa#1{\label{#1}\end{eqnarray}}
\def\eeqan{\end{eqnarray}}

\let\bar=\overbar

\def\Dslash{\not{\hbox{\kern-4pt $D$}}}
\def\dslash{\not{\hbox{\kern-2pt $\del$}}}

\def\msb{{\bar{\ssstyle M \kern -1pt S}}}

\usepackage{color}
\usepackage{lineno}
\usepackage{subfig}
\usepackage{hyperref}

\newcommand\pubnumber{PROC-CTD2020-37}

\newcommand\pubdate{\today}

\def\affiliation{
$^1$Cornell University, Ithaca, NY, USA 14853, \\
$^2$Fermi National Accelerator Laboratory, Batavia, IL, USA 60510\\
$^3$Princeton University, Princeton, NJ, USA 08544 \\
$^4$UC San Diego, La Jolla, CA, USA 92093 \\
$^5$University of Oregon, Eugene, OR, USA 97403
}

\newcommand{\conference}{Connecting the Dots Workshop (CTD 2020)\\
April 20-30, 2020}

\usepackage{fancyhdr}
\definecolor{mygrey}{RGB}{105,105,105}
\begin{document}
\newcommand{\mkFit}{\textsc{mkFit}\xspace}


\large
\begin{titlepage}
\pubblock

\vfill
\Title{Parallelizing the Unpacking and Clustering of Detector Data for Reconstruction of Charged Particle Tracks on Multi-core CPUs and Many-core GPUs}
\vfill

\Author{Giuseppe Cerati$^{2}$, Peter Elmer$^{3}$, Brian Gravelle$^{5}$, Matti Kortelainen$^{2}$, 
Vyacheslav Krutelyov$^{4}$, Steven Lantz$^{1}$, Mario Masciovecchio$^{4}$, Kevin McDermott$^{1}$, Boyana Norris$^{5}$, Allison Reinsvold Hall$^{2}$, Micheal Reid$^{1}$, Daniel Riley$^{1}$, 
Matev\v{z} Tadel$^{4}$, Peter Wittich$^{1}$, Bei Wang$^{3}$, Frank W\"urthwein$^{4}$, and Avraham Yagil$^{4}$ }
\Address{\affiliation}
\vfill

\begin{Abstract}
We present results from parallelizing the unpacking and clustering steps of the raw data from the silicon strip modules for reconstruction of charged particle tracks. Throughput is further improved by concurrently processing multiple events using nested OpenMP parallelism on CPU or CUDA streams on GPU. The new implementation along with earlier work in developing a parallelized and vectorized implementation of the combinatoric Kalman filter algorithm has enabled efficient global reconstruction of the entire event on modern computer architectures. We demonstrate the performance of the new implementation on Intel Xeon and NVIDIA GPU architectures.


\end{Abstract}
\vfill

\begin{Presented}
\conference
\end{Presented}
\vfill
\end{titlepage}
\def\thefootnote{\fnsymbol{footnote}}
\setcounter{footnote}{0}
%

\normalsize 


\section{Introduction}
\label{intro}
The reconstruction of charged particle trajectories (tracking) is a pivotal element of the reconstruction chain in Compact Muon Solenoid (CMS)~\cite{Chatrchyan:2008aa} as it measures the direction and momentum of charged particles, which is then also used as input for nearly all high-level reconstruction algorithms: vertex reconstruction, b-tagging, lepton reconstruction, jet isolation and missing transverse momentum reconstruction. Tracking by far is the most time consuming step of event construction and its time scales poorly with the detector occupancy. This brings a computing challenge to the upcoming upgrade of the accelerator from the Large Hadron Collider (HLC) to the High-Luminosity LHC (HL-HLC) where the instantaneous luminosity will increase by an order of magnitude due to the increased number of overlapping proton-proton collisions.

To address this challenge, the parallel Kalman filter tracking project \mkFit was established in 2014 with the goal of enabling efficient tracking on modern computing architectures. Over the last 6 years, we have made significant progress towards developing a parallelized and vectorized implementation of the combinatoric Kalman filter algorithm for tracking \cite{Cerati_2017,Cerati_2019,2019arXiv190611744C,2020arXiv200206295C}. This allows the efficient global reconstruction of the entire event within the projected online CPU budget. This also opens the possibility of deploying \mkFit into the CMS High Level Trigger (HLT), where the performance requirements are particularly strict. The current goal is to test the algorithm online in Run 3 of the LHC. 

Global reconstruction necessarily entails the unpacking and clustering of the hit information from all silicon strip tracker modules before the hots are processed by \mkFit. The current CMS HLT, on the other hand, performs hit and track reconstruction on demand, i.e., only for softwarer-selected regions of the detector. Therefore, we have recently begun to investigate how to implement the unpacking and clustering steps efficiently for the entire detector at once. This document highlights the latest development in enabling parallelization of unpacking and clustering on modern computing architectures. We start from a standalone version of the unpacker, which uses simulated raw data and calibration data, along with simplified versions of many related CMSSW classes. 

\section{Parallelization}
\label{para}
\begin{figure}[!htb]
  \centering
  \includegraphics[scale=0.5]{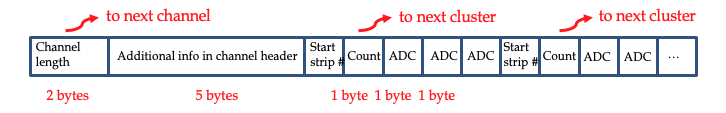}
  \caption{Raw data format from CMS strip tracker}
  \label{fig:SSTlayout}
\end{figure}

The CMS strip tracker data are organized by Front End Drivers (FEDs). Each FED consists of 96 optical channels with 256 strips per channel~\cite{Collaboration_2014}. In every CMS event, only the strips that are measured to be above a threshold are recorded. In zero suppression mode, the measured signal within a FED is stored like a variant of compressed sparse row (CSR) format, where the channel and strip numbers correspond to the row and column numbers (Figure ~\ref{fig:SSTlayout}). Besides the event-based strip tracker data, pre-measured calibration data, e.g., gain and noise values for each strip, must also be unpacked by FED channels. 

Data layout is paramount to performance, and therefore we choose the structure-of-array (SoA) layout. The SoA approach maximizes spatial locality on streaming access and thus improves sustained bandwidth on both modern CPUs and GPUs. The unpacking step transforms all event-based strip tracker data and pre-measured calibration data to the SoA format in order to provide optimal performance for the clustering step. Specifically, we construct a map of the channel locations and attributes on CPU. The construction can only be done sequentially, but can overlap with raw data transfer. We then unpack the data to SoA format, which can be done concurrently for each channel on a multicore CPU or GPU. Unpacking is the most time consuming step since it involves irregular data access pattern and is particularly costly on GPU. 

\begin{figure}[!htb]
  \centering
  \includegraphics[scale=0.4]{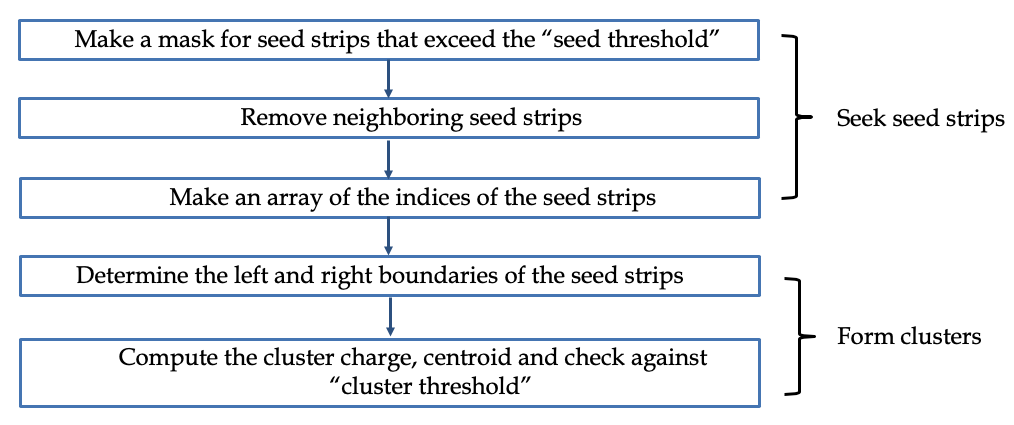}
  \caption{Parallel clustering algorithm workflow}
  \label{fig:workflow}
\end{figure}

The CMS clustering is based on the ``three thresholds" algorithm~\cite{Collaboration_2014}. First only strips with signal-to-noise ratio larger than the ``channel threshold" are recorded and we call these strips as active strips. Our parallel implementation is based on the fact that every candidate cluster will have at least one active strip with signal-to-noise ratio larger than the ``seed threshold", which we call seed strips. We start by building an array of indices of seed strips. We then form a cluster around a seed strip by determining the left and right boundaries, computing the cluster charge, and checking it against the ``cluster threshold". For each cluster that passes the ``three thresholds", we output its left and right boundaries, centroid and optionally ADC values. Figure~\ref{fig:workflow} shows the workflow of the parallel clustering algorithm. The seed seeking stage can be parallelized over all strips and the cluster forming stage can be parallelized over all seed strips. The input data for these tests is a simulated data sample of $t\bar{t}$ events with an average pileup per event of 70 using the Phase 1 CMS geometry in 2018 with realistic detector conditions. For this TTBar PU70 data sample, the strip number and the seed strip number are roughly 800,000 and 150,000, respectively. 

\section{Optimization}
\label{opt}
In order to maximize the hardware usage and improve the overall throughput for the standalone program, we have enabled concurrency in processing multiple events. On CPU, this is achieved with OpenMP nested parallelism by creating two levels of OpenMP parallel regions, one inside the other. The outer level is used to handle different events while an inner level is used to handle sets of strips within a single event. To maintain good memory locality in nested parallelism, it is important to ensure that threads are given affinity to specific cores. It is also important that memory is placed locally to the processor that accesses the data. On GPU, we use CUDA streams to launch multiple events concurrently on the same device. This not only maximizes GPU utilization, it also reduces data transfer overhead by overlapping communication with computation. We rely on a cached memory allocator developed in CMSSW~\cite{2020arXiv200404334B} based on one in the CUB library~\cite{cub} for GPU memory allocation. This has significantly reduced the memory allocation and deallocation overhead in processing multiple events. 

\section{Performance Results}
We evaluate the parallel performance of the standalone program on Tigergpu, the GPU cluster at research computing of Princeton University. Each node in Tigergpu is equipped with two 2.4 GHz Xeon Broadwell E5-2680 v4 processors on the host and four 1328 MHz NVIDIA P100 GPU as accelerators. For CPU results, each test uses all 28 cores with nested parallelism. For GPU results, we use OpenMP to stream multiple events concurrently on a single GPU. Table \ref{tab:perf} shows the wall-clock time and throughput results in running 840 events, where all events are copies of a single TTBar PU70 event. On CPU, the best performance is observed in running 14 events concurrently with 2 cores per event, while on GPU, running 2 events/streams concurrently results in the best performance. Note that the performance comparison is between running the code on a single GPU and on 28 CPU cores. 
\begin{table}[!htb]
    \begin{minipage}{.5\linewidth}
      \centering
          \caption*{2x2.4 GHz Xeon Broadwell E5-2680 v4}    
          \begin{tabular}{c|cc}
      \hline
      \hline
      NxM & time (s) & throughput (events/s) \\
      \hline
      1x28 &  1.7 & 492  \\
      2x14 &   2.1 & 400  \\
      4x7 & 1.67 & 502 \\
      7x4 & 1.57 & 532 \\
      14x2 & 1.50 & 560 \\
      28x1 & 2.10 & 400 \\
      \hline
      \hline
    \end{tabular}
        \end{minipage}%
    \begin{minipage}{.5\linewidth}
      \centering
             \caption*{1238MHz NVIDIA P100}    
      \begin{tabular}{c|cc}
      \hline
      \hline
      N &  time (s) & throughput (events/s)  \\
      \hline
      1 & 1.36 & 615 \\
      2 & 1.29 & 649 \\
      4 & 1.41 & 595 \\
      7 & 1.46 & 574 \\
      14 & 1.53 & 548 \\
      28 & 1.54 & 546 \\
      \hline
      \hline
    \end{tabular}
        \end{minipage} 
     \caption{Performance comparison in running 840 TTBar PU70 events. N is the events concurrency and M is the CPU parallelization within one event. The reported number for each test is the average of 10 trials.}
     \label{tab:perf}
     \end{table}

\section{Conclusions}
We have enabled parallelization of unpacking and clustering of CMS silicon strip detector data and demonstrated its performance on both multi-core CPUs and many-core GPUs. Overall, we observe that a single GPU (P100) outperforms two sockets Intel Broadwell CPU (2x14 cores 2.4 GHz Xeon Broadwell E5-2680 v4) by 24\% in average. Next, we will integrate the implementation with CMSSW and convert clusters to global hit coordinates in order to provide the necessary input for \mkFit. 


\Acknowledgements
This work is supported by the U.S. National Science Foundation, under grants PHY1520969, PHY1521042, PHY1520942 and PHY1624356, and under Cooperative Agreement OAC1836650, and by the U.S. Department
of Energy, Office of Science, Office of Advanced Scientific Computing Research, Scientific Discovery through Advanced Computing (SciDAC) program. This research used resources at research computing of Princeton University.





\bibliographystyle{unsrt}
\bibliography{references}

\begin{thebibliography}{1}

\bibitem{Chatrchyan:2008aa}
S.~Chatrchyan et~al.
\newblock {The CMS Experiment at the CERN LHC}.
\newblock {\em JINST}, 3:S08004, 2008.

\bibitem{Cerati_2017}
Giuseppe Cerati et~al.
\newblock Parallelized kalman-filter-based reconstruction of particle tracks on
  many-core processors and gpus.
\newblock {\em EPJ Web of Conferences}, 150:00006, 2017.

\bibitem{Cerati_2019}
Giuseppe Cerati et~al.
\newblock Parallelized and vectorized tracking using kalman filters with cms
  detector geometry and events.
\newblock {\em EPJ Web of Conferences}, 214:02002, 2019.

\bibitem{2019arXiv190611744C}
Giuseppe {Cerati} et~al.
\newblock {Speeding up Particle Track Reconstruction in the CMS Detector using
  a Vectorized and Parallelized Kalman Filter Algorithm}.
\newblock {\em arXiv e-prints}, page arXiv:1906.11744, June 2019.

\bibitem{2020arXiv200206295C}
Giuseppe {Cerati} et~al.
\newblock {Reconstruction of Charged Particle Tracks in Realistic Detector
  Geometry Using a Vectorized and Parallelized Kalman Filter Algorithm}.
\newblock {\em arXiv e-prints}, page arXiv:2002.06295, February 2020.

\bibitem{Collaboration_2014}
The~CMS Collaboration.
\newblock Description and performance of track and primary-vertex
  reconstruction with the {CMS} tracker.
\newblock {\em Journal of Instrumentation}, 9(10):P10009--P10009, oct 2014.

\bibitem{2020arXiv200404334B}
Andrea {Bocci}, David {Dagenhart}, Vincenzo {Innocente}, Christopher {Jones},
  Matti {Kortelainen}, Felice {Pantaleo}, and Marco {Rovere}.
\newblock {Bringing heterogeneity to the CMS software framework}.
\newblock {\em arXiv e-prints}, page arXiv:2004.04334, April 2020.

\bibitem{cub}
https://nvlabs.github.io/cub/.

\end{thebibliography}

\end{document}